\documentclass{LMCS}

\def\dOi{9(3:29)2013}
\lmcsheading%
{\dOi}
{1--12}
{}
{}
{Oct.~24, 2012}
{Sep.~30, 2013}
{}

\ACMCCS{[{\bf Theory of computation}]: }

\keywords{Alternating Turing machines, inductive and hyper-elementary
languages, arithmetical hierarchy, polynomial-time hierarchy}
\ACMCCS{[{\bf Theory of computation}]: Turing machines; Complexity classes; Higher order logic}
\amsclass{03D10,03D60,03D70}

\usepackage{hyperref,enumitem}

\usepackage{amsmath,amssymb,amscd,oldlfont}
\usepackage{proof}

\newcommand{\bnk}{\hbox{\scriptsize $\sqcup$}}

\newcommand{\save}[1]{}

\newcommand{\ul}[1]{\underline{#1}}

\newcommand{\cdlongarrow}[4]{\begin{CD} #1 @>#2 >#3 > #4\end{CD}}

\newcommand{\ara}[1]{{\stackrel{#1}{\longrightarrow}}} 
\newcommand{\uara}[1]{{\underset{#1}{\longrightarrow}}} 
\newcommand{\uarda}[1]{{\underset{#1}{\longtwoheadrightarrow}}} 
\newcommand{\longtwoheadrightarrow}{\longrightarrow\hspace{-1.2em}\rightarrow\hspace{.2em}}

\newcommand{\outt}[1]{}
\newcommand{\delete}[1]{}
\newcommand{\blank}[1]{}
\newcommand{\notedomission}[1]{\medskip\noindent{\bf TEXT OMITTED}\\[2mm]}

\newcommand{\bxit}[1]{\hbox{\it #1}}

\newcommand{\bxtt}[1]{\hbox{$\tt #1$}}

\newcommand{\calL}{\hbox{$\cal L$}}

\newcommand{\gra}{\hbox{$\alpha$}}

\newcommand{\grg}{\hbox{$\gamma$}}
\newcommand{\grd}{\hbox{$\delta$}}
\newcommand{\gre}{\hbox{$\varepsilon$}}

\newcommand{\grt}{\hbox{$\tau$}}

\newcommand{\grf}{\hbox{$\varphi$}}

\newcommand{\grG}{\hbox{$\Gamma$}}
\newcommand{\grD}{\hbox{$\Delta$}}

\newcommand{\grP}{\hbox{$\Pi$}}
\newcommand{\grS}{\hbox{$\Sigma$}}

\newcommand{\ra}{\rightarrow}

\newcommand{\rA}{\Rightarrow}  

\newcommand{\lng}{\langle}
\newcommand{\rng}{\rangle}

\newcommand{\ignore}[1]{}

\newcommand{\zero}{\rule{0mm}{3mm}}

\newcommand{\bc}{\begin{center}}
\newcommand{\ec}{\end{center}}
\newcommand{\beq}{\begin{equation}}
\newcommand{\eeq}{\end{equation}}

\newcommand{\be}{\begin{enumerate}}
\newcommand{\ee}{\end{enumerate}}
\newcommand{\bi}{\begin{itemize}}
\newcommand{\ei}{\end{itemize}}
\newcommand{\bd}{\begin{description}}
\newcommand{\ed}{\end{description}}
\newcommand{\beqn}{\begin{equation}}
\newcommand{\eeqn}{\end{equation}}

\newcommand{\beqna}{\begin{eqnarray}}
\newcommand{\eeqna}{\end{eqnarray}}
\newcommand{\beqnas}{\begin{eqnarray*}}
\newcommand{\eeqnas}{\end{eqnarray*}}
\newcommand{\beqnaa}{$$\begin{array}{rcll}}  
\newcommand{\eeqnaa}{\end{array}$$}  
\newcommand{\beqnana}{$$\begin{array}{lrcll}}  
\newcommand{\eeqnana}{\end{array}$$}  
\newcommand{\btbl}[1]{\begin{center}\begin{tabular}{#1}}
\newcommand{\etbl}{\end{tabular}\end{center}}
\newcommand{\beqnc}{$$\begin{array}{rclcl}}
\newcommand{\eeqnc}{\end{array}$$}
\newcommand{\fn}{\footnote}

\newtheorem{dclprop}{{\sc Proposition}} 
\newtheorem{dclprops}{{\sc Proposition}}[subsection] 
\newtheorem{dclbigthm}[dclprop]{THEOREM}

\makeatletter
\def\thmlabel#1{\@bsphack\if@filesw {\let\thepage\relax
\xdef\@gtempa{\write\@auxout{\string
\newlabel{#1}{{\@Roman{\@currentlabel}}{\thepage}}}}}\@gtempa
\if@nobreak \ifvmode\nobreak\fi\fi\fi\@esphack}
\makeatother

\newtheorem{dclthms}[dclprops]{{\sc Theorem}}   

\newtheorem{dcllems}[dclprops]{{\sc Lemma}} 
\newtheorem{dclsublem}[dclprop]{{\sc Sublemma}}

\newtheorem{dclcors}[dclprops]{{\sc Corollary}} 
\newtheorem{dcldfn}[dclprop]{{\sc Definition}}
\newtheorem{dcldfns}[dclprops]{{\sc Definition}}
\newtheorem{dclasss}[dclprops]{{\bf Assumption}}
\newtheorem{dclass}[dclprop]{{\bf Assumption}}

\newenvironment{dfn}{\medskip\begin{dcldfn}\sl}{\end{dcldfn}}

\newcommand{\bsl}{\begin{verse}\sl}
\newcommand{\esl}{\end{verse}}

\newtheorem{exs}[dclprop]{Examples}

\newtheorem{exxs}[dclprop]{Exercises}

\newenvironment{exercises-with-preamble}{\begin{exxs}\rm}{\end{exxs}}

\newcommand{\bthm}{\begin{thm}}
\newcommand{\ethm}{\end{thm}}
\newcommand{\bprop}{\begin{prop}}
\newcommand{\eprop}{\end{prop}}
\newcommand{\blem}{\begin{lem}}
\newcommand{\elem}{\end{lem}}
\newcommand{\bcor}{\begin{cor}}
\newcommand{\ecor}{\end{cor}}
\newcommand{\bdfn}{\begin{dfn}}
\newcommand{\edfn}{\end{dfn}}

\newcommand{\bz}{\begin{quote}\small}
\newcommand{\ez}{\end{quote}}

\newcommand{\einference}[2]  
  {\shortstack
      {$ #1 $\\ \mbox{}\\ $ #2 $}}

\newlength{\txtlth}
\newlength{\txtht}

\newcommand{\ttplus}{\bxtt{+}}
\newcommand{\ttminus}{\bxtt{-}}
\usepackage{times}

\begin{document}

\title[Alternation and Pi-1-1]{Alternating Turing machines for inductive languages}
\author[D.~Leivant]{Daniel Leivant}
\address{Indiana University}
\email{leivant@indiana.edu}  
\thanks{This research was also supported by LORIA Nancy and Universit\'{e} 
	Paris-Diderot}

\begin{abstract}
We show that alternating Turing machines, with a novel and natural
definition of acceptance,
accept precisely the inductive ($\grP_1^1$) languages.
Total alternating machines, that either accept or reject
each input, accept precisely the hyper-elementary ($\grD_1^1$) languages.
Moreover, bounding the permissible number of alternations yields
a characterization of the levels of the arithmetical hierarchy.
Notably, these results use simple finite computing devices, 
with finitary and discrete operational semantics, and
neither the results nor their proofs make any use of transfinite ordinals.

Our characterizations elucidate the analogy between the 
polynomial-time hierarchy and the arithmetical hierarchy, 
as well as between their respective limits,
namely polynomial-space and $\grP_1^1$.

\end{abstract}

\maketitle

\section{Introduction}

Inductive definitions via first-order positive operators constitute a broad
computation paradigm.
A fundamental result of computation theory, formulated in various
guises over the last century, 
identifies the languages obtained by such definitions with
those explicitly definable by $\grP_1^1$ formulas, that is where second
order quantification, over functions or relations, is restricted
to positive occurrences of $\forall$.
This central link was first discovered by
Suslin in 1916 for sets of real numbers \cite{Suslin17}. 
Kleene independently rediscovered the correspondence
for sets of natural numbers (and so for languages) \cite{Kleene55,Kleene55a}.
Spector formulated the basic notions more explicitly \cite{Spector61},
and Moschovakis, Barwise, and Gandy established the characterization for
near-arbitrary countable first-order structures in 1971. 
This characterization of $\Pi^1_1$ in terms of inductive definability
endows it with many of the structural
properties of the computably enumerable (RE) sets, and suggests
an analogy between computability based on finite processes,
captured by $\grS^0_1$,
and a generalized form of computability based on 
infinite processes.

Our aim here is to capture the full power of inductive definability 
by a novel and natural definition of acceptance
for alternating Turing machines.
This is unrelated to notions of ``infinite-time computations" 
that have been investigated repeatedly over the decades.

Alternation in computational and definitional processes is an idea
that has appeared and reappeared in various guises
over the last 50 odd years. Kleene's definition
of the arithmetical hierarchy in terms of quantifier alternation
was an early manifestation, extended by Kleene, Spector, Gandy and
others to the transfinite hyper-arithmetical hierarchy 
\cite{Rogers67,Barwise-admissible}.
An explicit link with alternation was discovered by Moschovakis 
\cite{Moschovakis71,Moschovakis74,Kolaitis-game}, who characterized  
the inductive sets by a game quantifier \cite[Theorem 5C2]{Moschovakis74}.
Harel and Kozen \cite{HarelK84} 
showed how this characterization can be expressed
in terms of an idealized programming language with random existential 
and universal assignments.

Alternating control made an entry into Computation Theory with the
definition, by Chandra, Kozen and Stockmeyer, of alternating Turing
machines \cite{ChandraKS81}, where existential and universal
variants of non-determinism mesh.
A state declared existential
accepts when some child-configuration accepts,
whereas a universal state accepts if all child-configurations
accept. A computation can thus alternate
between existential and universal phases.
The striking result of \cite{ChandraKS81}, which has become a classic and
has made its way into numerous textbooks, is that alternating
Turing machines elucidate a powerful and elegant interplay between time and 
space complexity.
Namely, for reasonable functions $f$ the languages
accepted by alternating Turing machines in time $O(f)$ are precisely
the languages accepted by deterministic machines
in space $O(f)$, and the languages accepted by alternating machines
in space $O(f)$ are those accepted by deterministic machines in time
$2^{O(f)}$.  
In particular, alternating polynomial time is precisely polynomial
space.  Moreover, when only fewer than $k$ alternations are allowed,
one obtains the $k$'th level of the polynomial time hierarchy.

We establish here a formal parallel between the logical and
the complexity-theoretic developments of alternation. 
Our point of departure is a
simple and natural modification of the definition of acceptance
by an alternating Turing machine, where acceptance by
a universal configuration $c$ will now refer to all configurations that end
the universal computation-phase spawned by $c$, rather than just to the
immediate children of $c$.
We prove that a language is accepted by
such a machine iff it is inductive ($\grP_1^1$).
Moreover, when up to $k$ alternations are allowed, 
we obtain the $k$'th levels of the arithmetical hierarchy.
Also, if a language $L$ is accepted by a machine which is total, in the
sense that every input is either accepted or rejected, then $L$ is
hyper-arithmetical ($\grD_1^1$).

Note that our machines are no different from traditional alternating
Turing machines: the difference lies only in the definition of acceptance.
In particular, no infinitary rules, such as game quantifiers or random
assignments, are used.
%
%
We thus obtain here a direct correspondence between 
$\grP_1^1$ and polynomial space, and between
the arithmetical hierarchy and the polynomial-time hierarchy. 
The two sides of this correspondence
are characterized by the same alternating Turing machines, 
but with a global (potentially infinitary) definition of acceptance 
for the former, and a local one for the latter.  



The author is grateful to Yiannis Moschovakis
for important comments on an early draft of this paper.

\section{Global semantics for alternating computations}

\subsection{Alternating Turing machines}

The following will be used as reserved symbols, which we posit
to occur only when explicitly referred to: $\bnk$ for the blank symbol,
\ttplus\ for the cursor-forward command, and \ttminus\ for cursor-backward.
We consider primarily single-tape machines.
Given a finite alphabet \grS, 
an {\em alternating Turing machine (ATM) over \grS} is a device
$M$ consisting of
\be
\item
Disjoint finite sets $E$ (existential states) and $U$ (universal states).
Elements of $Q = E \cup U$ are the {\em states.}
\item
An element $s_0 \in Q$, dubbed the {\em start state}.
\item
A finite alphabet $\grG\supseteq \grS\cup \{\bnk\}$ (the {\em machine alphabet}).
\item
A relation $\grd \subseteq (Q \times \grG) \; \times \; (A \times Q)$,
where $A = \grG \cup \{\ttminus,\ttplus\}$ is the set of {\em actions.}\fn{We follow
here the convention whereby Turing machines either move their cursor or overwrite it,
but not both.}
\zero \grd\ may be construed as a multi-valued function, with
domain $Q \times \grG$ and co-domain $A \times Q$.
We write
$\cdlongarrow{q}{\gamma (a)}{M}{q'}$ \quad
for \quad $(q,\grg,a,q') \in \grd$,
and omit the subscript $M$ when in no danger of confusion.
\ee

\noindent A {\em configuration (cfg)} (of $M$) is 
a tuple $(q,u,\grg,v)$ with 
$q \in Q$, $u,v \in \grG^*$, and $\grg\in \grG$.
A cfg is said to be {\em existential} or {\em universal} 
according to the state therein.
The definition of a yield relation $c \rA c'$ between configurations
is defined as usual; that is, it is generated inductively by the conditions:\fn{Note that
inductive definitions posit implicitly an exclusivity condition, so the
``only if" direction is not needed.}

\bi
\item
If \quad $\begin{CD} q @>\gamma (+) >M > q'\end{CD}$ \quad then \quad
$(q,u,\grg,\grt v) \rA (q',u\grg,\grt,v)$ \quad and \quad
$(q,u,\grg,\gre) \rA (q',u\grg,\bnk,\varepsilon)$;\fn{We write $\varepsilon$ 
for the empty string.}
\item
If \quad $\begin{CD} q @>\gamma (-) >M > q'\end{CD}$ \quad then \quad
$(q,u\grt ,\grg,v) \rA (q',u,\grt,\grg v)$ \quad and \quad
$(q,\gre,\grg,v) \rA (q',\gre, \grg,v)$  (i.e.\ the cursor does not move); and
\item
If \quad $\begin{CD} q @>\gamma (\tau) >M > q'\end{CD}$ \quad then \quad
$(q,u,\grg,v) \rA (q',u,\grt,v)$.
\ei

Following \cite{Kozen06} we dispense here with accepting
and rejecting states:
when no transition applies to a universal
cfg then it has no children, and so the condition for
acceptance is satisfied vacuously. Dually, a dead-end existential
cfg is rejecting.
For brevity we also write configurations $(q,u,\grg,v)$ as a pairs $(q,w)$, where
the understanding is that $w$ is a ``cursored string" $u \ul{\grg} v$.




%

\subsection{Acceptance and rejection}

The {\em computation tree of $M$ for cfg $c$} is
a finitely-branching (but potentially infinite) tree $T_M(c)$
of cfg-occurrences $\lng \gra,c\rng$, \gra\ being the node-address and
$c$ the cfg,
where the children of $\lng \gra,c\rng$
are $\lng i\gra,c_i\rng$ with $c_i$ the $i$-th cfg $c'$ such that
$c \rA c'$ (under some fixed ordering of the transition rules of \grd).

We write $c \uara{\exists} c'$ when $c \rA c'$ and $c$ is existential,
$c \uarda{\exists} c'$ if $c \uara{\exists}^* c'$ and $c'$ is universal.
(As usual,
$\uara{\exists}^*$ is the reflexive and transitive closure of $\uara{\exists}$.)
In other words, the universal cfg $c'$ can be reached from the cfg
$c$ by successive applications of the yield relation $\rA$, where all intermediate 
states are existential.

The definitions of $c \uara{\forall} c'$ and $c \uarda{\forall} c'$
are similar.
We call a cfg $c'$ as above, for either
$\uarda{\exists}$ or $\uarda{\forall}$,
an {\em alternation-pivot (for $c$)}.
 
The set \bxit{AC} of {\em accepted configurations} is generated inductively
by the following closure conditions:
\be
\item
If $c$ is existential and $c'\in AC$ for some $c'$
such that $c \uarda{\exists} c'$,
then $c \in AC$.
\item
If $c$ is universal and $c'\in AC$ for all $c'$ such that 
$c \uarda{\forall} c'$, then $c \in AC$.
\ee
If $S$ is any set of configurations, 
we write $\bxit{CC}[S]$ for the conjunction of the conditions above for $S$.
That is,
\be
\item
If $c$ is existential and $c'\in S$ for some $c'$
such that $c \uarda{\exists} c'$, then $c \in S$.
\item
If $c$ is universal and $c'\in S$ for all $c'$ such that
$c \uarda{\forall} c'$, then $c \in S$.
\ee
Thus, \bxit{AC} is generated by the closure conditions
$\bxit{CC}[\bxit{AC}]$.
Note that $\bxit{CC}[S]$ is a $\grP^0_2$ formula.
For instance, (2) can be expressed as
$$
\forall \; \text{cfg} \;\; c \quad
((\forall \; \text{traces witnessing a relation} \; 
		c \uarda{\forall} c') \quad c' \in \bxit{S})
	\quad \ra c \in \bxit{S}
$$

Thus, the set \bxit{AC} of accepted configurations is explicitly definable
as the set of configurations $c$ satisfying the $\grP_1^1$ formula
$$
\forall S \; (\bxit{CC}[S] \ra \; c \in S)
$$
Similarly, the set \bxit{RC} of {\em rejected configurations} is 
generated inductively by closure conditions dual to the ones above:
\be
\item
If $c$ is existential and $c'\in RC$ for all $c'$ such that
$c \uarda{\exists} c'$, then $c \in RC$.
\item
If $c$ is universal and $c'\in RC$ for some $c'$ such that
$c \uarda{\forall} c'$, then $c \in RC$.
\ee
Again, \bxit{RC} is explicitly definable by a $\grP_1^1$ formula.

\medskip

We say that a state is {\em dead-end} if no transition rule
applies to it.
A universal dead-end state is an {\em accept-state}, and an 
existential dead-end state a {\em reject-state.}

\medskip

The {\em initial configuration} of the machine $M$ for input $w$ is
$\lng s_0, \gre,\bnk,w\rng$.  $M$ {\em accepts an input string $w$} if the initial
cfg for $w$ is in the set \bxit{AC} of accepted configurations, as 
defined above.  Dually,
$M$ {\em rejects $w$} if that cfg is in \bxit{RC}.
For example, if $M$ has only universal states, then no computation tree 
can have an alternation-pivot, and so every $w$ is accepted.
The computation tree for $w$ may well have leaves, that is dead-end configurations,
but since here these are all universal
configurations with no children, they are accepted.
Dually, if $M$ has only existential states, then no input can be accepted.
These examples are merely consequences of our choice to represent acceptance and
rejection by dead-end universal and existential configurations, respectively.
For example, a usual non-deterministic Turing machine can be obtained simply 
by considering each accept-state as a universal state with no applicable 
transition rule.

The {\em language accepted by an ATM $M$} is
$$
\calL(M) = \{ w \in \grS^* \mid M \hbox{ accepts } w \}
$$
and the {\em language rejected by $M$} is
$$
\bar{\calL}(M) = \{ w \in \grS^* \mid M \hbox{ rejects } w \}
$$
It is easy to see that $\calL(M) \cap \bar{\calL}(M) = \emptyset$.
Our definitions of acceptance and rejection of configurations conform to the
local closure conditions of acceptance (and rejection) of usual ATMs, as we point
out in the next Proposition.  
However, those conditions cannot be used to {\em define} acceptance
and rejection, because we allow infinite computation trees.

\bprop\label{prop:truth-value-closure}
Let $M$ be an ATM, $T$ a computation tree of $M$ for input $w$.
If $c$ is a cfg in the tree, with children $c_1 \ldots c_m$, then
\be
\item
If $c$ is existential, then $c$ is accepted
iff some $c_i$ is accepted, and $c$ is rejected iff all $c_i$'s are rejected.
\item
If $c$ is universal, then $c$ is accepted
iff all $c_i$'s are accepted, and $c$ is rejected iff some $c_i$ is rejected.
\ee
\eprop
\proof
Let $c$ be existential.
If $c$ is an accepted cfg, i.e.
$c \uarda{\exists} c'$ for some accept-state $c'$,
then $c_i \uarda{\exists}^* c'$ for some $c_i$, since $c$ itself is existential.
If that $c_i$ is existential, then it is accepted, by definition;
and if it is not, then $c_i=c'$, which is accepted by assumption.

Conversely, suppose that some $c_i$ is accepted.
If $c_i$ is universal, then $c \uarda{\exists} c'$, and so
$c$ is accepted, by definition of acceptance.
If $c_i$ is existential, then
there must be an accepted $c'$ such that $c_i \uarda{\exists}^* c'$;
but then $c \uarda{\exists}^* c'$, so $c$ is accepted.

Other cases are proved similarly.
\qed

\subsection{Divergence and totality}

An ATM may well neither accept nor reject an input string $w$.
For example, if the computation tree of $M$ for a given
input $w$ has infinitely many alternation-pivots along
each computation-trace (a situation that we can engineer fairly easily),
then $M$ neither accepts nor rejects that input.
Indeed, the empty set satisfies the closure conditions for acceptance
of $w$, as well as the closure conditions for rejection.

We say that an ATM $M$ is {\em total} if every input is either accepted or 
rejected by $M$. Let us identify a simple condition that guarantees totality.
We say that a computation tree is {\em alternation well-founded}
if no branch has infinitely many alternation-pivots.
An ATM is {\em alternation well-founded} if all its computations are 
alternation well-founded.

\bprop\label{prop:awf-is-total}
If an ATM is alternation well-founded then it is total.
\eprop
\proof
We prove the contra-positive:
if a cfg $c$ is neither accepted nor rejected,
then the computation tree $T$ that it spawns has a branch with infinitely
many alternation-pivots.

Suppose $c$ is universal. Since $c$ is not accepted, we must have $c \uarda{\forall} c'$
for some alternation-pivot $c'$ which is not accepted.  And since
$c$ is also not rejected, all of its alternation-pivots,
and in particular $c'$, are not rejected.   If $c$ is existential, a
dual argument shows that $c\uarda{\forall} c'$ for some alternation-pivot $c'$ which is neither
accepted nor rejected. 

Iterating the argument we obtain a branch with an infinite sequence
$c_0=c, c_1 = c' , \ldots$ of successive alternation-pivots, 
all of which are neither accepted nor rejected.
\qed

\medskip

The converse of Proposition \ref{prop:awf-is-total} fails.
Indeed, it is easy to construct a total ATM that is not alternation
well-founded, by inserting innocuous computation traces with
infinitely many alternation-pivots, with no impact
on the acceptance or rejection of the input.
See the proof of Proposition \ref{prop:one-sided} below.

\subsection{Duality and one-sidedness}

The {\em dual} of an ATM $M$ is the machine $\bar{M}$
whose transition relation is that of $M$, but with the sets of universal
and existential states interchanged, that is
with $M$'s sets $U$ and $E$ as the sets of existential and universal
states, respectively.

Directly from the definitions we have

\bprop\label{prop:duality}
Let $\bar{M}$ be the dual of $M$.
Then $\calL(\bar{M}) = \bar{\calL}(M)$, whence also
$\bar{\calL}(\bar{M}) = \calL(M)$.
\qed
\eprop

A machine $M$ is {\em one-sided} if it either has no accepted
configurations, or no rejected configurations.

\bprop\label{prop:one-sided}
For every machine $M$ there are one-sided machines
$M^+$ and $M^-$ such that $\calL(M)= \calL(M^+)$,
and $\bar{\calL}(M)= \bar{\calL}(M^-)$.
\eprop
\proof
The proof is analogous to the conversion of a deterministic TM to 
a TM that diverges for any input it does not accept.

Let $M^+$ be obtained from $M$ by expanding its
transition relation as follows.
Using auxiliary states and transitions, we add
for every existential state a transition into an auxiliary universal
state that starts an infinite trace (using auxiliary states) of alternation-pivots.
That is, we create a fresh alternation-pivot following each existential cfg,
where that alternation-pivot is neither accepted nor rejected.
Each state accepted in $M$ is accepted in $M^+$, because no existential configuration
is loosing any pivot by the modification.
And if a state is accepted in $M^+$, then it is accepted in $M$, because
the set $A$ of configurations of $M^+$ that consists just of the accepting configurations
of $M$ satisfies the closure conditions \bxit{AC} for $M^+$, 
and therefore contains the set of configurations accepted by $M^+$ (which is the
minimal such set).

But $M^+$ has no rejected configurations: existential configurations cannot be rejected
because they have an alternation-pivot, namely the one introduced by the definition of
$M^+$, which is not rejected.  And then universal configurations cannot be rejected,
because all their alternation-pivots, which are existential, are non-rejected.

The construction of $M^-$ is dual.
\qed

\subsection{The Arithmetical Hierarchy}


We say that an ATM $M$ {\em is $\grS_k$} if its initial state is existential,
and for every $w \in \grS^*$, all branches of the 
computation-tree for $w$ have $\leq k$ alternation-pivots.
The definition of $\grP_k$ machines is similar, but with a
universal initial state.
Here again we posit that the existential states of $\grP_1$ machines have
no applicable transition rules.

%
%

\bthm\label{thm:arith-hie}
Let $k \geq 1$.
A language is $\grS_k^0$ ($\grP_k^0$) iff it is accepted by a
$\grS_k$ ($\grP_k$, respectively) ATM.
\ethm
\proof
The proof is by induction on $k$.
For the base case $\grS^0_1$, let $L$ be a language
defined by a $\grS^0_1$ formula, that is
$$
L \quad = \qquad \{ x \in \grS^* \mid \grf[x] \}
$$
where
$$
\grf[x] \; \equiv \;\; \exists w_1 , \ldots , w_r\; \grf_0[\vec{w},x]
$$
with $\grf_0$ a bounded formula,
i.e.\ with all quantifiers bounded (under the substring relation).
Define a $\grS_1$ machine $M$ that accepts $L$, as follows.
$M$ branches existentially to choose a string
$w = w_1 \# \cdots \# w_r$,
then proceeds to check deterministically that $\grf_0[w_1 \ldots w_r]$.
(We classify the states for that deterministic process
to be universal, so that dead-end states are accepted.)

For the converse, note first that in a $\grS_1$ computation tree the universal configuration
are all accepted, since they have no pivots. So acceptance by a 
a $\grS_1$ machine $M$ is definable by the $\grS^0_1$ formula that states,
for input $w$, the existence of a finite tree of configurations, related
by the rules of $M$, with the
initial configuration for $w$ as root,
of which the internal nodes are existential and the leaves are universal.

For the base case $\grP^0_1$, suppose $L$ is defined by a $\grP^0_1$ formula 
$$
\grf[x] \; \equiv \; \forall w_1 \ldots w_r\; \grf_0[w_1, \ldots, w_r,x]
$$
Define a $\grP_1$ machine $M$ that accepts $L$, as follows.
$M$ generates strings $w_1 \# \cdots \# w_r$ in successive lexicographic
order.  After each such choice $M$ branches universally to the next 
string as well as to a deterministic module
that accepts $x$ iff $\grf_0[\vec{w},x]$ for the current value 
of $w_1 \# \cdots \# w_r$.

Conversely, if $L = \calL(M)$ where $M$ is an $\grP_1$ machine, then
$L$ is definable by a formula that states that for all (finite) computation
traces, the trace's last configuration is not existential (i.e.\ rejected).

The induction step generalizes the induction basis:
The properties above are proved for level $k\!+\!1$ of the Arithmetical
Hierarchy by referring to sub-computations at level $k$,
rather than to deterministic sub-computations.
\qed

\section{Alternation and inductive languages}

\subsection{Accepted languages are 
	inductive}\label{subsec:accepted-is-inductive}

Fix an alphabet \grS.
Consider formulas over the vocabulary (i.e.\ similarity type) with
an identifier for each letter in \grS\ as well as for the
empty-string, a binary function-identifier for concatenation, and a binary
relation for the substring relation.

\bprop\label{prop:dfns-inductive}
The following conditions are equivalent for a language $L \subseteq \grS^*$.
\bd
\item[I1]
\zero $L$ is  defined by a formula of the form $\forall f \; \grf[w,f]$,
where \grf\ is first-order and $f$ ranges over $\grS^* \ra \grS^*$.
\item[I2]
\zero $L$ is defined by a formula of the form 
$\forall f \; \exists x \; \grf_0[w,f,x]$, where $\grf_0$ is a bounded formula,
i.e.\ with each quantifier restricted to substrings of some string.
\item[I3]
\zero $L$ is defined by a formula of the form $\forall f \; \exists x \;
	\grf_0[w,\bar{f}(x),x]$, where 
$\bar{f}(x)$ abbreviates the string $f(0)\# \cdots \# f(|x|)$
(with $\#$ a fresh symbol, used as a textual separator).
\item[I4]
\zero $L$ is defined by a formula of the form
$\forall S \;\exists x\; \forall y  \; \grf_0[w,f,x,y]$, where
$S$ ranges over subsets of $\grS^*$.
\ed
\eprop
\proof
I1 implies I2 by the Kuratowski-Tarski algorithm \cite{KuratowskiT31}.
I2 implies I3 by the boundedness of $\grf_0$. I1 implies I4 by
an interpretation of functions by relations (and hence sets, since we
are talking about languages), and I3 and I4 each implies I1 trivially.
\qed

\medskip

Note that the use of a set quantifier in I4 necessitates
an alternation of first-order quantifiers, which is not needed in I1.
This is essential: without the presence of the first-order universal
quantifier $\forall y$ we get Kreisel's {\em strict-$\Pi_1^1$} formulas,
which are no more expressive than $\grS_1^0$ \cite{Kreisel60,Kreisel68}.

A language $L\subseteq \grS^*$ is {\em inductive} ($\grP^1_1$) when it
satisfies the equivalent conditions of Proposition \ref{prop:dfns-inductive}
(see e.g.\ \cite{Kleene52}).

Recall that our definition above of acceptance by an ATM 
refers to the set \bxit{AC} of accepted configurations, which
is $\grP_1^1$ definable.  We therefore have:

\bprop\label{prop:inductive-accepted}
Every language accepted by an ATM is inductive.
\eprop


\subsection{Inductive languages are accepted}

\bprop\label{prop:acepted-inductive}
Every inductive language is accepted by an ATM.
\eprop
\proof
We refer to characterization (I3) of $\grP^1_1$ languages.
As usual, $\grS^n$ stands for the set of strings over \grS\ of
length $n$.
Let $L$ be a language defined by
$$
	\forall f \exists x \; \grf_0[w,\bar{f}(x),x]
$$
which we write momentarily as
$$
\forall f\; \exists \,x \; \grf_0[w,\, z_0 \# \cdot\cdot\cdot \# z_n,\,x]
$$
where $n = |x|$ and $z_i = f(i)$.
This is equivalent to the following infinite formula (where, as usual, $\grS^n$
is the set of strings of length $n$).

\newcommand{\dollar}{\#}

\beqn\label{eq:pi11-inf}
\begin{array}{lllll}
\grf_0[w,\gre,\gre]\\
\quad \vee\; \forall z_0 \;\; (\exists x\in \hbox{{\small $\Sigma^1$}} \; \grf_0[w,z_0,x])\\
        \quad\quad \vee \; \forall z_1  
		\; (\exists x\in \hbox{{\small $\Sigma^2$}} \; 
			\grf_0[w,z_0\, \dollar\,z_1,x])\\
        \quad\quad\quad \vee \; \forall z_2  
		\; (\exists x\in \hbox{{\small $\Sigma^3$}} \; 
			\grf_0[w,z_0\, \dollar\,z_1\, \dollar\,z_2,x])\\
        \quad\quad\quad\quad\quad \vee \; \forall z_3  
		\; (\exists x\in \hbox{{\small $\Sigma^4$}} \; 
		\grf_0[w,z_0\, \dollar\,z_1\, \dollar\,z_2\, \dollar\,z_3,x])\\
        \quad\quad\quad\quad\quad\quad\quad\quad \vee \; \cdot\,\cdot\,\cdot
\end{array}
\eeqn
We use here infinitary formulas for informal expository purpose;
compare \cite{Moschovakis74,Moschovakis71}. 

Formula (\ref{eq:pi11-inf}) is captured by an ATM which, on input $w$,
\be
\item
checks deterministically $\grf_0[w,\gre,\gre]$; if this fails,
\item
chooses by universal nondeterminism a value $z_0$;\fn{Recall from the
introduction that such a choice, for our finitely-branching machine,
involves a computation tree with an infinite branch.}
\item
for each such choice for $z_0$, branches by existential 
nondeterminism to
\be
\item guess (by existential nondeterminism) an $x\in \grS$,
then check (deterministically) $\grf_0[w,z_0,x]$;  if this fails
\item
choose by universal nondeterminism a $z_1$;
\item
etc.\qed
\ee
\ee

%

\medskip

\noindent Combining Propositions \ref{prop:inductive-accepted} and
\ref{prop:acepted-inductive}
we conclude:

\medskip

\bthm\label{thm:inductive=acceptable=awkacceptable}
$L$ is inductive iff it is accepted by an ATM.
\ethm

\section{Total machines and hyper-arithmetical languages}



A basic result of computation theory is the characterization of
decidable languages in terms of semi-decidability:

\bthm\label{thm:RE-coRE} A language $L \subseteq \grS^*$ 
is accepted by a Turing machine that terminates for all input
iff both $L$ and its complement are accepted by a Turing machine.
\ethm


The analog of Theorem \ref{thm:RE-coRE} is

\bthm\label{thm:total=accepted+coaccepted}
A language $L \subseteq \grS^*$ is accepted by a total ATM iff both $L$ and its 
complement $\bar{L} = \grS^* - L$ are accepted by an ATM.
\ethm

The forward implication of the Theorem is easy:
If a language $L$ is accepted by a total ATM $M$ then
the dual machine $\bar{M}$ accepts $\bar{L}$, by
by Proposition \ref{prop:duality}.

Towards proving below the converse implication,
assume that $L= \calL(M_0)$ and $\bar{L}= \calL(N)$.
By Proposition \ref{prop:one-sided} we may assume that neither 
machine has rejected configurations. Thus $\bar{L}$ is rejected by the
machine $M_1 = \bar{N}$, which has no accepted cfg.
We wish to construct out of $M_0$ and $M_1$ a total machine $M$
that accepts $L$ and rejects $\bar{L}$.
A naive emulation of the
standard proof of Theorem \ref{thm:RE-coRE}
would swap control between $M_0$ and $M_1$ after each computation step.
That is, $M$ is defined as a two-tape machine, whose states are 
tuples $\lng q_0,q_1,j\rng$,
with $q_i$ a state of $M_i$,
and where $j\in \{0,1\}$ indicates which
machine is to make a move.
The type of $\lng q_0,q_1,j\rng$ (existential or universal) is 
the type of $q_j$.
Since $M_1$ has no accepted cfg, 
a cfg $c$ of $M$ would be accepted when its $M_0$ 
component is accepted by $M_0$;
and since $M_0$ has no rejected configurations,
$c$ would be rejected in $M$ if its $M_1$ component is rejected by $M_1$.

However, the construction above does not work for our ATMs, due to the global
definition of acceptance.
Consider a universal cfg $c_0$ of $M_0$,
which is accepted in $M_0$ because it has no pivots.
When $c_0$ is coupled in $M$ with a universal cfg $c_1$ of $M_1$,
the combined cfg may spawn a computation tree with pivots 
of $M_1$, whose $M_0$-component is not accepted in $M_0$.
The combined cfg is not accepted then in $M$, even though
$c_0$ is accepted in $M_0$.

We consider instead a merge of $M_0$ and $M_1$ where control swap
from a universal phase of $M_0$ to $M_1$ is
delayed until that phase has ended, and dually for an existential phase of $M_1$.
%

Note that for simple Turing machines (deterministic or nondeterministic)
phases coincide with computation steps, since no universal
configurations are present.

More precisely, we posit, without loss of generality, 
that $M_0$ and $M_1$ are single-tape ATM's over 
a common input alphabet $\grS$, and
using a common extended alphabet $\grG \subset \grS \cup \{\bnk\}$.
The combined machine $M$ is then defined as follows. 

\bi
\item
\zero $M$ is a two-tape ATM, whose states of interest are tuples 
$\lng q_0,q_1,j\rng$, 
with $q_i$ a state of $M_i$, $(i=0,1)$.
The type of $\lng q_0,q_1,j\rng$ (existential or universal) is 
the type of $q_j$ (in $M_j$).
\item
In addition, $M$ has auxiliary states and (deterministic) transitions that
pre-process its computation by copying the
input into the second tape, reinitializing the cursor positions, 
and passing control to a state
$\lng s_0,s_1,0\rng$, where $s_i$ is the initial state of $M_i$.
\item
For $\grg,\grd\in \grG,\; \gra \in \{+,-\}\cup \grG$,\\
if \quad $q_0 \; \ara{\gamma  (\alpha)} \; p_0$ \quad is a rule of $M_0$ then
\bi
\item 
If both $q_0$ and $p_0$ are universal, then 
$$
\cdlongarrow{\lng q_0, q_1, 0 \rng}
	{\gamma,\delta (\alpha, \delta)}{}{\lng p_0,q_1,0\rng}
$$
i.e.\ on reading \grg\ on the first tape,
and \grd\ on the second, $M$ performs action \gra\ on component 0
of the cfg, action \grd\ (i.e.\ no-op) on component 1,
and leaves control to component 0.
\item
Otherwise, i.e.\ if at least one of $q_0$, $p_0$ is existential, then 
$$
\cdlongarrow{\lng q_0, q_1, 0 \rng} 
	{\gamma,\delta (\alpha, \delta)}{}
	{\lng p_0,q_1,1\rng}
$$
\ei
\item
If \quad $q_1 \; \ara{\gamma (\alpha)} \; p_1$ is a rule of $M_1$, then
\bi
\item
If both $q_1$ and $p_1$ are existential, then 
$$
\cdlongarrow{\lng q_0, q_1, 1 \rng}{\gamma,\delta (\alpha, \delta)}{} 
	{\lng q_0,p_1,1\rng}
$$
\item
Otherwise, i.e.\ if at least one of $q_1,p_1$ is universal, then
$$
\cdlongarrow{\lng q_0, q_1, 1 \rng}
	{\gamma,\delta (\alpha, \delta)} {}
	{\lng q_0,p_1,0\rng}
$$
\ei
\ei


\bprop\label{prop:combined-machine}
Assume that no string is both accepted by $M_0$ and rejected by $M_1$.
Then $M$ accepts every string accepted by $M_0$.
\eprop

\proof
We prove that if $M_0$ accepts a cfg $(q_0,u_0)$ then, 
for every non-rejected cfg $(q_1,u_1)$  of $M_1$,
$M$ accepts $(\lng q_0,q_1,0\rng, \lng u_0,u_1\rng)$.
If $M_0$ accepts $u$, then (by assumption) $M_1$ does not reject it,
and so the Proposition follows by considering the cfg $(\lng,s_0,s_1,0\rng, \lng u,u\rng)$.

%
%

Define the set $A$ of $M_0$-configurations by
$$
\begin{array}{ll}
A = \{ (q_0,u_0) \mid & (\lng q_0,q_1,0\rng, \lng u_0,u_1\rng) \quad
	\hbox{is accepted in $M$} \\
	& \qquad \hbox{for all
	non-rejected configurations $(q_1,u_1)$ of $M_1$}
	\; \}
\end{array}
$$
We show that $A$ satisfies the closure conditions defining
the set of configurations accepted by $M_0$.
\bi
\item
\zero {\bf The existential closure condition:}
Suppose that $(q_0,u_0) \uarda{\exists} (p_0,w_0)$, where $(p_0,w_0) \in A$,
and the reduction sequence is of length $n \geq 1$.\fn{$n=0$ is excluded,
since by definition of  $\uarda{\exists}$ the state $q_0$ is existential
and $p_0$ is universal.}
We prove that $(q_0,u_0) \in A$ by induction on $n$.
Let $(q_0,u_0) \uara{\exists} (r_0,v_0) \uarda{\exists} (p_0,w_0)$.
Note, first, that we must have $(r_0,v_0) \in A$:
if $n=1$ then $(r_0,v_0) = (p_0,w_0) \in A$,
and if $n>1$ then $(r_0,v_0) \in A$ by IH.

Towards proving that $(q_0,u_0) \in A$ let 
$(q_1,u_1)$ be a non-rejected cfg of $M_1$.
We have
$$
(\lng q_0,q_1,0\rng, \,\lng u_0,u_1\rng) 
	\; \uara{\exists} \; 
	(\lng r_0,q_1,1\rng, \,\lng v_0,u_1\rng)
$$
We show that $(\lng r_0,q_1,1\rng, \,\lng v_0,u_1\rng)$
is accepted in $M$, whence so is
$(\lng q_0,q_1,0\rng, \,\lng u_0,u_1\rng)$.

We have the following cases.
\bi
\item
\zero $q_1$ is universal.
Each $M_1$-cfg $(r_1,v_1)$ such that
$(q_1,u_1) \uara{\forall} (r_1,v_1)$ must be non-rejected, since 
$(q_1,u_1)$ is non-rejected.
We have
$$
(\lng r_0,q_1,1\rng, \,\lng v_0,u_1\rng) 
	\; \uara{\forall} \; 
	(\lng r_0,r_1,0\rng, \,\lng v_0,v_1\rng)
$$
and the latter cfg is accepted,
since $(r_0,v_0) \in A$ and $(r_1,v_1)$ is non-rejected.
It follows that $(\lng r_0,q_1,1\rng, \,\lng v_0,u_1\rng)$ is accepted
in $M$.

\item
\zero $q_1$ is existential.
Since $(q_1,u_1)$ is non-rejected, it follows
that $(q_1,u_1) \uarda{\exists} (r_1,v_1)$ for some
non-rejected cfg $(r_1,v_1)$ of $M_1$.
By definition of $M$, we have
$$
(\lng r_0,q_1,1\rng, \,\lng v_0,u_1\rng)
        \; \uarda{\exists} \;
        (\lng r_0,r_1,0\rng, \,\lng v_0,v_1\rng)
$$
The latter cfg is accepted, 
since $(r_0,v_0) \in A$, and $(r_1,v_1)$ is non-rejected.
It follows that $(\lng r_0,q_1,1\rng, \,\lng v_0,u_1\rng)$ is accepted
in $M$.
\ei

We have thus shown that if $\;(q_0,u_0) \uarda{\exists} (p_0,w_0)$, 
where $(p_0,w_0) \in A$, then
$(q_0,u_0) \in A$.

\item
\zero {\bf The universal closure condition:}
Suppose that for all $(p_0,w_0)$, 
if $(q_0,u_0) \uarda{\forall} (p_0,w_0)$ then
$(p_0,w_0) \in A$. 
Towards showing that $(q_0,u_0) \in A$, let
$(q_1,u_1)$ be a non-rejected cfg of $M_1$.


By definition of $M$, if $(\lng q_0,q_1,0\rng, \,\lng u_0,u_1\rng)
        \; \uarda{\forall} \; C$, where $C$ is a cfg of $M$,
then 
$C = (\lng p_0,q_1,1\rng,\,\lng w_0,u_1\rng)$,
where $(q_0,u_0) \uarda{\forall} (p_0,w_0)$.

We show that $(\lng p_0,q_1,1\rng, \,\lng w_0,u_1\rng)$
is accepted in $M$ for each such $(p_0,w_0)$,
implying that $(\lng q_0,q_1,0\rng, \,\lng u_0,u_1\rng)$
is accepted.

We have the following cases.
\bi
\item
\zero $q_1$ is universal.
Suppose $(q_1,u_1) \uara{\forall} (p_1,v_1)$.
Then
$$
(\lng p_0,q_1,1\rng, \,\lng w_0,u_1\rng) 
 	\; \uara{\forall} \; 
 	(\lng p_0,p_1,0\rng, \,\lng w_0,v_1\rng)
$$
The cfg $(p_1,v_1)$ must be non-rejected, since 
$(q_1,u_1)$ is non-rejected. Since $(p_0,w_0) \in A$, it follows that
$(\lng p_0,p_1,0\rng, \,\lng w_0,v_1\rng)$ is accepted.
This being the case for every $(p_1,v_1)$ as above,
we conclude that $(\lng p_0,q_1,1\rng, \,\lng w_0,u_1\rng)$ is accepted.


\item
\zero $q_1$ is existential.
Since $(q_1,u_1)$ is non-rejected, it follows
that $(q_1,u_1) \uarda{\exists} (p_1,v_1)$ for some
non-rejected configuration $(p_1,v_1)$ of $M_1$.
By definition of $M$, we have
$$
(\lng p_0,q_1,1\rng, \,\lng w_0,u_1\rng)
         \; \uarda{\exists} \;
         (\lng p_0,p_1,0\rng, \,\lng w_0,v_1\rng)
$$
The latter configuration is accepted in $M$,
since $(p_0,w_0) \in A$ and $(p_1,v_1)$ is non-rejected.
It follows that $(\lng p_0,q_1,1\rng, \,\lng w_0,u_1\rng)$ is also accepted,
\ei
We have thus shown that if $(p_0,w_0) \in A$
whenever $\;(q_0,u_0) \uarda{\forall} (p_0,w_0)$,
then $(q_0,u_0) \in A$, that is $A$
satisfies the universal closure condition for acceptance in $M_0$.
\ei

\noindent Since $A$ satisfies both the existential and the universal
closure conditions for acceptance in $M_0$,
it follows that $A$ contains every accepting cfg of $M_0$,
proving the Proposition.
\qed

\medskip

\noindent
{\bf Proof of Theorem \ref{thm:total=accepted+coaccepted} --- Concluded.}
We have noted already the forward implication.
We show that if $L$ and $\bar{L}$ are accepted by ATM's,
then $L$ is accepted by a total ATM.

Let $L = \calL(M_0)$
and $\bar{L} = \calL(N)$, and refer to the machines $M_1$ and $M$ of the discussion above.
By Proposition \ref{prop:combined-machine},
$M$ accepts every string accepted by $M_0$.

An argument dual to that in the proof of Proposition \ref{prop:combined-machine} 
shows that $M$ rejects every cfg $(\lng q_0,q_1\rng,\lng u_0,u_1\rng)$,
where $M_1$ rejects $(q_1,u_1)$ and $M_0$ does not accept $(q_0,u_0)$.
In particular, assuming $M_1$ rejects an input string $u$,
$M$ rejects $(\lng t_0,s_1,1 \rng, \lng v,u\rng)$ whenever 
$(t_0,v)$ is a non-accepted configuration of $M_0$.

A small extra step is needed to account for the fact that $M_0$, and not $M_1$,
has the initial control in $M$.
Posit, without loss of generality, that the initial state $s_0$ of $M_0$
is existential and deterministic (i.e.\ at most one transition applies).
Since $M_0$ does not accept $u$, 
we must have $(s_0,u) \uara{\exists} (t_0,v)$
where $(t_0,v)$ is a non-accepted cfg.
But then the unique initial transition of $M$ (past the initialization phase) is
$$
(\lng s_0,s_1,0\rng,\lng u,u\rng) \uara{\exists} (\lng t_0,s_1,1\rng,\lng v,u\rng)
$$
Since $M_1$ rejects $(s_1,u)$ and $M_0$ does not accept $(t_0,v)$,
$M$ must reject $(\lng t_0,s_1,1\rng,\lng v,u\rng)$, as noted above, and therefore
must also reject $(\lng s_0,s_1,0\rng,\lng u,u\rng)$.

In conclusion, $M$ accepts every string accepted by $M_0$,
and rejects every string rejected by $M_1$.
So $M$ is a total machine that accepts $L$ and rejects $\bar{L}$.
\qed

\section{Conclusion}

The combined use of existential and universal nondeterminism
has been of interest primarily in Computational
Complexity theory, but has not been considered thus far 
as a tool in the foundations of computing.  
This is because the semantics of acceptance has been
defined ``locally", that is in terms of the relation
between computational configurations and their immediate descendants.
That semantics implies that acceptance (and rejection) are witnessed by
finite computation trees, and thus cannot lead us beyond the semi-decidable
(RE) languages.  Viewed from another angle, the closure properties
involved are $\grP^0_1$, and so the accepted languages are defined
by strict-$\Pi^1_1$ formulas 
(see \S\ref{subsec:accepted-is-inductive} above).

We showed here that a very natural alternative semantics for
universal nondeterminism changes the picture radically,
as the languages accepted are precisely the $\grP^1_1$ ones.
This further illustrates the foundational analogy
between alternation in feasible time with local semantics, which
yields PSpace as a limit of the PTime Hierarchy (starting with PTime and NP), 
and alternation for arbitrary computations with global semantics, which
yields $\grP^1_1$ as a limit of the arithmetical hierarchy 
(starting with $\grS_1^0$).

Generalized models of computation that go beyond computability have
been studied extensively, of course. The novelty of the
approach here is that it refers to the very same hardware as traditional
Turing machines (albeit with both modes of nondeterminism), but redefines
the notion of acceptance, in a way that remains consistent with the
underlying, intuitive, intent.

The ability to refer to both computational complexity
and higher recursion theory using the same machine models has the
potential of suggesting analogies between results, and thereby
transfer of results.
We believe that this will provide insights and additional
machine-based proofs for Higher Recursion Theory.  

The approach developed here seems to also break with past works
in this area in that it dispenses with transfinite recurrence and induction
over Kleene's constructive ordinals, and does not use any transfinite
stage-comparison technique.
Instead, the proofs use inductive definitions directly.


Directly dealing with inductive definitions, without calibrating
them by ordinals provides, in fact, a closer analogy with finite
computing.
Computation traces of machines and of programs are construed
intuitively as finite objects,
without direct reference to the natural numbers, either as clocking
computation steps or as codes for computational objects.
Wit the frequent use of ``structural induction" and ``structural recurrence."
It is, therefore, natural to expect that higher-order computation traces
can be studied directly, without a detour through transfinite clocking by
constructive ordinals.  The proof of Theorem 
\ref{thm:total=accepted+coaccepted} achieves precisely that.


\bibliographystyle{plain}
\bibliography{x}
\end{document}